\documentstyle[twocolumn,aps]{revtex}
\begin{document}

\newcommand{\beq}{\begin{equation}}
\newcommand{\eeq}{\end{equation}}
\newcommand{\bea}{\begin{eqnarray}}
\newcommand{\eea}{\end{eqnarray}}
\newcommand{\ba}{\begin{array}}
\newcommand{\ea}{\end{array}}
\newcommand{\om}{(\omega )}
\newcommand{\bef}{\begin{figure}}
\newcommand{\eef}{\end{figure}}
\newcommand{\leg}[1]{\caption{\protect\rm{\protect\footnotesize{#1}}}}

\newcommand{\ew}[1]{\langle{#1}\rangle}
\newcommand{\be}[1]{\mid\!{#1}\!\mid}
\newcommand{\no}{\nonumber}
\newcommand{\etal}{{\em et~al }}
\newcommand{\geff}{g_{\mbox{\it{\scriptsize{eff}}}}}
\newcommand{\da}[1]{{#1}^\dagger}
\newcommand{\cf}{{\it cf.\/}\ }
\newcommand{\ie}{{\it i.e.\/}\ }
\newcommand{\eg}{{\it e.g.\/}\ }

\title{FAQ about the  ``contextual objectivity"  point of view.}
\author{Philippe Grangier}
\address{Laboratoire Charles Fabry de l'Institut d'Optique, 
F-91403 Orsay, France}

\maketitle

\begin{abstract}

We discuss some Frequently Asked Questions about the ``contextual objectivity" 
point of view on quantum mechanics introduced in two previous preprints \cite{ph1,ph2}.


\end{abstract}

\section{Introduction}

In a previous preprint \cite{ph1}, we introduced and discussed a ``physical"
(as opposed to mathematical)  definition of  
a quantum state\footnote{Throughout this paper ``state" means ``pure state",
unless it is specified as ``mixed state". To avoid confusion,
pure states will also be called ``modalities" \cite{ph2}.}, 
that reads in the following way : 

{\bf The quantum state of a physical system is defined by the values of a 
complete\footnote{The set of quantities is
complete in the sense that the value of any other quantity which satisfies
the same criteria is a function of the set values.}
set of physical quantities, which can be predicted with certainty and measured
repeatedly without perturbing in any way the system}. 

As discussed in detail in  ref. \cite{ph1}, 
this definition is in full agreement with the usual formalism of QM.
It is also implies that some ``objectivity" can be attached to the quantum state,
because the quantum state is defined
from a fully predictable course of events, that is independent of the observer. 
In \cite{ph2} we tried to exploit this definition,
together with some ideas about the system dimensionality,
to propose a new axiomatic approach to QM, that is an attempt
to spell out how the ``quantum reality" is related with the ``macroscopic reality". 
While \cite{ph1} is a straightforward rewriting of usual QM, 
\cite{ph2} is much more tentative
and its goal is mostly to stimulate some thinking. 
We recommend to the reader to
have a look at least at \cite{ph1} before reading the FAQ below.

\section{Frequently Asked Questions (and answers)}


Q: What is contextual objectivity ?

A: It is an attempt to reformulate quantum mechanics (QM) in a more physical
(as opposed to mathematical) way. The ultimate goal of such a reformulation, 
that is evoked in \cite{ph2}, would be to explain why QM is the way it is. 
In particular, the Hilbert space structure should be deduced, not postulated.

It is worth emphasizing that contextual objectivity is {\it not} an alternative theory to
Quantum Mechanics, but it is rather an alternative way to present it: 
for all practical purposes, the predictions
drawn from our approach will be the same as the ones of standard QM, only the point
of view will be slightly different. So the difference is not in the 
prediction, but in the so-called ``interpretation" of QM. 
For instance, we claim that our point of view
is ``realistic", though it is certainly not a revival of ``local realism". We fully acknowledge
that QM is non-separable, but we emphasize that quantum non-separability is {\it not} an ``action at
a distance" (not even an ``influence at a distance"; see below).
We insist also on the uniqueness of the macroscopic world, at variance with the 
so-called ``many worlds" point of view (see also below).
\\
\\
Q: The quantum state cannot be an objective property of the system, 
because if you are given the system it is not possible to recover the state.

A: You should be given not only the system, but also the set of relevant observables
(\ie the ``context", or in usual terms the measurement basis). 
Then the state can obviously be recovered with certainty. 
\\
\\
Q: Does that mean that the set of relevant observables is ``an intrinsic part of the reality" 
(cf Bohr's answer to the EPR argument) ?

A: In some sense yes. As explained in \cite{ph2}, a very  specific quantum 
feature is the existence of  ``non-exclusive modalities" (in usual terms, non-orthogonal pure states): 
one cannot recover a state among a set of non-exclusive modalities, but it is quite possible
to recover it among a set of exclusive ones (in usual terms, orthogonal ones). 
This is why the set of relevant observables
is needed together with the system, and why we speak about ``contextual" objectivity. 
We note that the appropriate set of relevant observables is also observer-independant.
\\
\\
Q: How to explain the EPR ``paradox" ?

A: In an EPR state, the initial quantum state is a state of the  particles pair,
and the fully predictible quantities are the results of Bell measurements,
that are {\it joint} measurements on both particles. 
On the other hand, the states for each particle are undefined (they have no ``reality"). 
When Alice performs a measurement,
the state is redefined on her side. Given her measurement result, 
and assuming that she knew the initial entangled state,
Alice can infer Bob's state. On Bob's side, nothing changed :
there was no pure state before, and there is no pure state after,
until Alice informs Bob about what she measured.
For doing that she needs classical transmissions.
Therefore, Alice's measurement does not ``act upon" Bob's particle in any sense.
We note that Bob is also free to make a measurement on 
his side. If he does, it can be checked afterwards that his result is
compatible with (but not determined by) Alice's prediction. 

With respect to Bell's inequalities, one should notice that for an EPR state 
the strong correlations between measurements on the subsystems are 
due to global properties, while the properties of
each subsystem are completely random. 
{\it Such a situation is totally non-classical,
because classically correlations must be ``mediated" by properties
of the subsystems}. So Bell's hypothesis contradict QM
because of the failure of the ``local reality" (or ``separability") assumption, 
that states that  in order to explain the correlations, 
there should exist a property (a ``local hidden variable") 
that describes the polarization on each side\footnote{
Mathematically this is expressed by the dependance of the measurement
result $A(\vec a, \lambda) = \pm 1$ on both the adjustable parameter $\vec a$
and the hidden variable $\lambda$. Classically, if there is nothing like $\lambda$,
then $A(\vec a) = \pm 1$ is purely random, and the correlations
should vanish, while this is not the case quantum mechanically. This emphasizes
again that an essential hypothesis for Bell's inequalities is ``local reality".}. 
As said above, this is not the case in QM, but again there is no ``action at a distance",
not even any ``influence at a distance". 

The ultimate lesson from the EPR argument and Bell's inequalities
is that {\it classical physics is unable to manage global properties
of a system, that are not mediated by individual properties of the
subsystems, while QM can perfectly do so}\footnote{In case this sentence sounds
too ``holistic" to the reader, let us remind that in our approach, the
quantum state is always ``embedded" in a classical environment.}.
On the other hand, both classical and quantum correlations
are due to ``common causes in the past evolution", 
and thus obey relativistic causality. 
\\
\\
Q: Why not to say that a pure quantum state is a ``state of knowledge" ? 

A: An implicit consequence of the wording ``state of knowledge" is that
such a state should be contingent (\ie observer-dependant), and that it should be associated
with an ``ignorance" of something. This is indeed true for a mixed state,
but this is {\bf not} true for a pure state, that is observer-independant, and
that is not associated with the ``ignorance" of anything (\ie, there is no
hidden variables). Therefore, one might use a wording like ``objective state of knowledge",
but this is not very clear, and this is why we prefer to say that  a state is ``real"
in the contextual objectivity point of view. Actually, as said above, 
a pure state appears to be
real and objective in the usual sense as long as one carries out measurements
within the specified complete set of commuting observables,
following the time evolution of the system. 
\\
\\
Q: How to deal with the problem of the boundary between quantum and classical ``realities" ?

A: The discussion in \cite{ph2}
does not directly answer the question of the quantum-classical boundary,
but rather makes it irrelevant, by claiming that 
the {\it true basic postulate of QM is
the existence of both a continuous classical world and a quantized quantum world}.
The structure of QM is then a consequence from the need to connect them together. 
As said above, this line of reasoning is an
attempt to stimulate thinking rather than a proof, and we summarize below
its main points.

A physical quantity is defined as an ensemble of possible measurements,
that are connected between themselves by ``geometrical" transformations
that we call  ``knob transformations" (they may be standard geometrical
transformations, such as rotations of  a Stern-Gerlach magnet...). 
This definition of physical quantities, that cannot be avoided in our 
opinion, is essentially
classical~: it cannot be quantum, since it actually defines
the parameters that will be used to measure the state of the quantum system\footnote{
In an extreme view, what is required is the (continuous) ``spacetime" in which the experiment
is carried out, that is {\it not} a quantum space-time (at least, not in usual QM).}. 
More precisely, we will assume that the ``knob transformations" have
the structure of a {\it non-commutative continuous group}\footnote{
In a more precise definition, physical quantities will
be associated with the infinitesimal generators of that group.}.
On the other hand, the quantum system is intrinsically {\it quantized}:
though it may be in an infinite number
of (non-exclusive) modalities, we {\it postulate}
that for a given ``context" (\ie complete set of commuting observables),
there is only a discrete number of exclusive modalities, that is
a property of the system (its dimension). Not surprisingly, quantization 
is the main feature of the quantum system.

Given the two concepts of physical quantities and of a system,
the most naive classical approach is to identify each physical quantity with
the numbers given by the measurements, and to attribute ``reality" to these numbers.
EPR themselves realized that this definition of ``reality" was too
restrictive, and proposed instead their definition based upon
predictability and reproducibility; 
this is just the idea that we use as our definition of a quantum state. 
But as soon as this is done, 
it appears that ``reality" based upon predictability and reproducibility
cannot be attributed simultaneously
to all physical quantities : this is simply incompatible with 
the structure of the physical quantities described above.

One wants indeed to hold simultaneously 
that the physical quantities are defined by measurements, depending
on continuous parameters, while the possible measurement results
(the ``exclusive modalities") are quantized. 
How QM is able to make these two requirements fit together is discussed in \cite{ph2}
(see also \cite{res}).
We reach then the conclusion that what is ``real" at the macroscopic level is 
the definition of the physical quantities (\ie of the possible measurements), and 
what is ``real" at the quantum level (of the measured system)
is the quantum state. These two ``realities"
are fully compatible - they are actually the only ones that can
connect the experimental definition of a physical quantity and the measurement results 
in a consistent way.
\\
\\
Q: What about the ``many-worlds" interpretation ?

A: The so-called ``many-worlds" interpretation of QM is 
a possible (the only possible ?)
alternative to our point of view. The difference
between the two approaches is clear : while our approach 
is built to take into account that the measurement of
$S_z$ on a superposition $(|+\rangle + |-\rangle)/\sqrt{2}$ gives only
{\bf one} result, the ``many-worlds" approach claims that it gives
{\bf two} (totally equivalent) results. For consistency 
with classical reality, it adds that our mind is made is
such a way that it ``follows" only one of these possibilities. 
Rather that introducing our mind at that point (whose's mind actually ?),
we consider more fruitful to assume that physical reality 
is uniquely defined, within the framework of contextual objectivity.


\begin{references}
\bibitem{res}

For the sake of completeness here is a short summary :

- For a given ``knob settings" of the measurement apparatus,
there exist $N$ distinguishable quantum states $\{b_i\}$, 
that are called ``exclusive modalities". 
The value of $N$, called the dimension, is a characteristic 
property of a given quantum system.
It is {\it not} assumed that the $\{b_i\}$ are rays in
an Hilbert space.

- Different knob settings are related between themselves 
by transformations $g$ that
have the structure of a continuous group ${\cal G_{K}}$. 
We introduce the representation of the group ${\cal G_{K}}$
by $N \times N$ unitary matrices, and we denote as  
$\tilde\Sigma_g$ the  unitary matrix corresponding to the group element $g$. 

- If the system is known to be in the state $b_i$ from the set $\{b_i\}$,
the probability that it is found in state $b_j'$ from the set $\{b_j'\}$ 
corresponding to another knob settings obtained by the knob transformation $g$ is : 
\beq
p_{i,j} = Trace(P_i \;\;  \tilde\Sigma_g \; \; P_j  \; \; \tilde\Sigma_g^\dagger)
\eeq
where the $P_i$ are $N \times N$ orthogonal projectors, 
and $\tilde\Sigma_g$ is the unitary matrix
corresponding to $g$. 

The Hilbert space structure can be deduced from the above statements,
where the basic mathematical entities are the  $N \times N$ unitary
matrices $\tilde\Sigma_g$ that represent ${\cal G_{K}}$.
\end{references}
\end{document}